\documentclass[a4paper,11pt]{article}

\usepackage{jheppub}
\usepackage{comment}

\newcommand{\TEW}{T_\text{EW}}
\newcommand{\FaI}{F_{\alpha I}}

\newcommand{\Mp}{M_{\text{Pl}}}

\newcommand{\paren}[1]{\left( #1 \right)}
\newcommand{\Higgs}{H}
\newcommand{\GB}{\chi}
\newcommand{\sgt}{\phi}

\title{Dark matter, singlet extensions of the \boldmath{$\nu$}MSM, and symmetries}

\author{Kyle Allison}
\affiliation{Rudolf Peierls Centre for Theoretical Physics, University of Oxford,\\ 1 Keble Road, Oxford OX1 3NP, United Kingdom}
\emailAdd{k.allison1@physics.ox.ac.uk}

\abstract{We consider an extension of the $\nu$MSM in which sterile neutrino masses originate from the VEV of a Higgs singlet $\phi$ and dark matter is produced through the decays of $\phi$ rather than through active-sterile neutrino mixing.
This model, which we refer to as the $\nu$NMSM, can readily satisfy or escape the constraints on warm dark matter from the Lyman-$\alpha$ forest and other small scale structure. However, it requires a particular hierarchy of Majorana masses and Yukawa couplings without an obvious origin.
We show that the hierarchical parameters of the $\nu$NMSM can arise from symmetries broken at or near the Planck scale for two specific examples of this model: one in which $\phi$ helps stabilize the electroweak vacuum through a scalar threshold effect and one in which $\phi$ is a light inflaton. Both examples require a complex $\phi$ and have several experimental signatures that are distinct from the $\nu$MSM. These signatures include additional dark radiation that is relativistic at both primordial nucleosynthesis and CMB decoupling and, for the former, a large invisible branching ratio of the Higgs.}

\arxivnumber{1210.6852}

\begin{document}
\maketitle
\flushbottom

\section{Introduction}
\label{sec:intro}
The $\nu$MSM~\cite{Boy09} is an extension of the Standard Model (SM) that attempts to explain all observed phenomena beyond the SM using only three sterile neutrinos with Majorana masses below the electroweak scale.
In the $\nu$MSM, one sterile neutrino, $N_1$, is responsible for dark matter~\cite{Asa05} while two additional sterile neutrinos, $N_2$ and $N_3$, are responsible for baryon asymmetry production~\cite{Sha05}. Moreover, the Higgs boson with a non-minimal coupling to gravity is responsible for inflation~\cite{Bez08}.

Although a detailed study of the $\nu$MSM (see~\cite{Can12b} for a recent update) shows that this minimal model can explain most of the observed phenomena beyond the SM, there are several
indications that an extension of the $\nu$MSM, such as by a Higgs singlet, may be necessary:
\begin{itemize}
\item \emph{Lyman-$\alpha$ forest bound}: The Lyman-$\alpha$ forest~\cite{Boy09c} and other small scale structure~\cite{Mir07,Pol11} impose strong constraints on the non-resonant production of warm dark matter in the $\nu$MSM when combined with X/$\gamma$-ray limits~\cite{Wat12}. Several solutions to this problem have been proposed, including a resonant production of dark matter from a large lepton asymmetry~\cite{Sha08} and a dilution of dark matter from a late entropy release~\cite{Sha06b,Sha07a,Bez10b,Nem12}. Generating a sufficiently large lepton asymmetry requires an inverted neutrino hierarchy as well as a high level of fine-tuning or the use of an approximate Planck-scale symmetry and non-renormalizable operators~\cite{Roy10}, while generating a sufficiently large entropy dilution requires some new physics beyond the $\nu$MSM~\cite{Sha06b,Sha07a,Bez10b,Nem12}. An attractive alternative to these scenarios uses the decays of a Higgs singlet, whose vacuum expectation value (VEV) provides an origin for the Majorana masses of the sterile neutrinos, to give a primordial production of dark matter~\cite{Sha06,Ani09,Bez10,Kus06,Kus08}.
\item \emph{Electroweak vacuum stability}: For a Higgs mass in the range $m_h \simeq 125$--126~GeV~\cite{ATL12,CMS12}, the Higgs potential develops an instability below the Planck scale unless the top mass is about 2$\sigma$ below its central value; for its central value, an instabililty develops at $10^9$--$10^{10}$~GeV~\cite{Eli11}. While more precise measurements of the top mass may lower its central value and relieve this tension, it has been shown that, if necessary, the addition of a Higgs singlet below the instability scale can stabilize the potential through its contribution to the renormalization group evolution of the Higgs quartic coupling~\cite{Che12,Leb12} or through a tree-level scalar threshold effect~\cite{Leb12,Eli12}.
\item \emph{Higgs inflation}: There has been some discussion about unitarity violation and the self-consistency of Higgs inflation with a non-minimal coupling $\xi \Higgs^\dag \Higgs R$, where $R$ is the scalar curvature and $\xi \sim 10^4$ (see~\cite{Bez11} and references therein). In brief, this model of Higgs inflation violates unitarity at the scale $\Lambda_0 \sim \Mp / \xi$ when expanding about a small background Higgs field. Although the scale of unitarity violation is raised to $\Mp / \sqrt{\xi}$ when expanding about the large background Higgs field during inflation~\cite{Bez11}, if the theory is eventually embedded into a more complete one that is valid up to the Planck scale then new physics is expected to appear at $\Mp / \xi$~\cite{Giu11}. Several solutions that do not abandon the minimality of the model have been proposed, including non-renormalizable Higgs interactions that accompany the non-minimal coupling and restore unitarity~\cite{Ler10} as well as strong coupling in graviton exchange processes that only break unitarity perturbatively~\cite{Ler12}. However, it has not yet been shown that these scenarios can actually be realized~\cite{Ler12}. Alternatively, an extension of the $\nu$MSM by a Higgs singlet can ``unitarize'' Higgs inflation~\cite{Giu11}\footnote{In \cite{Ler12}, it is argued that this is not a true completion of Higgs inflation but rather an induced gravity inflation model added onto the SM.} or provide a workable scenario with the singlet as the inflaton~\cite{Sha06,Ani09,Bez10}.
\end{itemize}
The fact that a Higgs singlet can both provide an origin for the Majorana masses of the sterile neutrinos and allow a simple dark matter production mechanism that, unlike the non-resonant production of dark matter in the $\nu$MSM, is consistent with the Lyman-$\alpha$ forest bound is a strong motivation for considering singlet extensions of the $\nu$MSM (e.g.~\cite{Sha06,Ani09,Bez10,Kus06,Kus08}). It is then natural to ask whether such extensions can also address the issues with Higgs inflation, as in~\cite{Sha06,Ani09,Bez10}, or help stabilize the electroweak vacuum, if necessary.

These singlet extensions of the $\nu$MSM, like the original model, require a particular hierarchy of Majorana masses and Yukawa couplings without an obvious origin. An important open question for these extensions is whether it is possible for such structure to come from an underlying symmetry. In the context of the $\nu$MSM, models employing a U(1) flavour symmetry~\cite{Sha07,Mer11b}, discrete flavour symmetries~\cite{Ara12}, the split seesaw mechanism~\cite{Kus10,Adu11}, and the Froggatt-Nielsen mechanism~\cite{Mer11,Bar11,Bar12} have been suggested for producing a hierarchical pattern of Majorana masses and Yukawa couplings. Similar techniques should also be able to produce the necessary pattern of masses and couplings in singlet extensions, but this has not been demonstrated explicitly.

In this paper, we consider extensions of the $\nu$MSM by a Higgs singlet $\sgt$ that address some of the model's possible phenomenological problems and demonstrate how underlying symmetries can give the necessary pattern of Majorana masses and Yukawa couplings in these extensions.
In particular, our starting point is a generic model in which the decays of $\sgt$ allow for primordial dark matter production that is consistent with the Lyman-$\alpha$ forest bound and in which the VEV of $\phi$ provides an origin for the Majorana masses of the sterile neutrinos.
We then construct symmetries broken at or near the Planck scale that can produce the hierarchy of parameters for two specific examples of this model: one in which $\sgt$ helps stabilize the electroweak vacuum through a scalar threshold effect~\cite{Eli12} and one in which $\sgt$ is the inflaton~\cite{Sha06,Ani09,Bez10}. Both examples require a complex $\sgt$ to be realized with underlying symmetries and have several experimental signatures that are distinct from the $\nu$MSM.

The paper is organized as follows. In section~\ref{sec:vMSM}, we review the constraints on the $\nu$MSM and primordial dark matter production from a Higgs singlet. In section~\ref{sec:sym}, we develop symmetries broken at or near the Planck scale that can produce the required pattern of Majorana masses and Yukawa couplings for two examples of this model. Section~\ref{sec:conc} gives the conclusions.

\section{The \boldmath{$\nu$}MSM and dark matter production from a Higgs singlet}
\label{sec:vMSM}
In this section, we first review the constraints on the $\nu$MSM and motivate the scenario of dark matter production from a Higgs singlet. We then discuss the constraints on dark matter production from a Higgs singlet.

\subsection{The $\nu$MSM}
The Lagrangian of the $\nu$MSM is given by
\begin{equation}
\label{eq:lagrangian}
\mathcal{L} = {\mathcal{L}_\text{SM}} + {\bar N_I}i{\partial _\mu }{\gamma ^\mu }{N_I} - {F_{\alpha I}}{\bar L_\alpha }{N_I}\Higgs - \frac{{{M_{IJ}}}}{2} \bar N_I^c{N_J} + {\text{h.c.}},
\end{equation}
where $\mathcal{L}_\text{SM}$ is the SM Lagrangian, $N_I$ ($I = 1,2,3$) are the sterile neutrinos, $L_\alpha$ ($\alpha = e,\mu,\tau$) are the lepton doublets, $\Higgs$ and $\sgt$ are the Higgs doublet and singlet, respectively, $\FaI$ are the Yukawa couplings for neutrinos, and $M_{IJ}$ are the Majorana masses for the sterile neutrinos. One of the striking features of the $\nu$MSM is the highly constrained and hierarchical pattern of parameters required for successful baryogenesis and dark matter production. These constraints are often best expressed not in the basis $N_I$ of \eqref{eq:lagrangian} but in the basis of the \emph{physical} mass eigenstates $N_I^m$ with masses $M_I$ and Yukawa couplings $\tilde{F}_{\alpha I}$. The two bases are related by the unitary transformation given in~\cite{Sha07}.

First, consider the constraints on $N_2^m$ and $N_3^m$. The oscillations between $N_2^m$ and $N_3^m$ above $\TEW$ produce a lepton asymmetry in the active neutrinos that is converted into a baryon asymmetry by sphalerons~\cite{Sha05}.\footnote{$\TEW \simeq 140$~GeV for a Higgs mass $m_h \simeq 125$~GeV~\cite{Can10}.} $N_2^m$ and $N_3^m$ cannot enter thermal equilibrium at temperatures much above $\TEW$ or else the lepton asymmetry produced in their oscillations is wiped out, giving the constraint~\cite{Sha07}
\begin{equation}
\label{eq:F2}
F_2 \lesssim 1.2\times 10^{-6},
\end{equation}
where $F_I^2 \equiv \paren{F^{\dag} F}_{II}$ and, by convention, $F_2$ is taken to be larger than $F_3$ with $\epsilon \equiv F_3/F_2 \leq 1$. Similarly, masses $M_2,M_3 \ll \TEW$ are required so that lepton number violating processes are negligible for $T \gtrsim \TEW$; masses satisfying
\begin{equation}
\label{eq:Mless}
M_2, M_3 \lesssim 20\text{~GeV}
\end{equation}
are generally considered acceptable~\cite{Sha05,Can10}. Meanwhile, effective baryon asymmetry production requires $M_2,M_3 \simeq M$ to be highly degenerate with a mass difference $\Delta M \equiv M_3 - M_2 \ll M$~\cite{Sha05}. The baryon asymmetry produced can be expressed as a function of $F_2, \epsilon, M, \Delta M$, and the neutrino hierarchy. Since active neutrino masses are generated via the seesaw mechanism, one of $F_2, \epsilon,$ and $M$ (typically $F_2$) can be expressed in terms of the others with the relation~\cite{Sha07}
\begin{equation}
\label{eq:seesaw}
\Delta m_\text{atm} \simeq \frac{\kappa v^2 \epsilon F_2^2}{2M},
\end{equation}
where $\Delta m_\text{atm} \simeq 0.05$~eV, $v = 246$~GeV, and $\kappa = 1 (2)$ for the inverted (normal) hierarchy. Analytic expressions for the baryon asymmetry are given in~\cite{Sha05,Asa10} while a numerical study has been carried out in~\cite{Can10}. The allowed range of each parameter individually is~\cite{Can10}
\begin{gather}
\label{eq:M} M \gtrsim \text{140 MeV},\\
\label{eq:deltaM} 10^{-3} \text{~eV} \lesssim \Delta M \lesssim \text{MeV},\\
\label{eq:epsilon} 10^{-4} \lesssim \epsilon \leq 1,
\end{gather}
though the combination must produce the observed asymmetry $n_B/s \simeq \paren{\text{8.4--8.9}} \times 10^{-11}$~\cite{Sha09}. Note that the lower bound \eqref{eq:M} comes from demanding that $N_2^m$ and $N_3^m$ decay before Big Bang nucleosynthesis (BBN)~\cite{Sha08b,Ruc12}\footnote{Recent work~\cite{Gor12,Can12} suggests this bound can been strengthened to $M \gtrsim 1.4$~GeV in the RP scenario, discussed later.} and that a significant amount of parameter space for $M \lesssim 500$~MeV is ruled out by the CERN PS191 experiment and other accelerator bounds~\cite{Can10,Ruc12}. For the parameter space allowed by \eqref{eq:M}--\eqref{eq:epsilon}, a lower bound on $F_2$ is approximately
\begin{equation}
F_2 \gtrsim 3 \times 10^{-8}.
\end{equation}

Now consider the constraints on the dark matter candidate $N_1^m$. Mixing with active neutrinos below $\TEW$ allows the 1-loop decay $N_1^m \rightarrow \nu^m \gamma$ with width~\cite{Pal82,Boy09}
\begin{equation}
\begin{aligned}
\label{eq:gammaN1}
\Gamma_{N_1^m \rightarrow \nu^m \gamma} &= \frac{9\alpha G_F^2}{1024\pi^4}\sin^2\paren{2\theta_1}M_1^5 \\
&\simeq 5.5 \times 10^{-22} \theta_1^2 \paren{\frac{M_1}{\text{keV}}}^5\text{s}^{-1},
\end{aligned}
\end{equation}
where $\theta_1^2 = v^2 \tilde{F}_1^2 /\paren{ 2 M_1^2}$ and $\tilde{F}_1^2$ is evaluated with~\cite{Sha07}
\begin{equation}
\label{eq:F1tilde}
\tilde{F}_{\alpha 1} \sim F_{\alpha 1} + \frac{M_{12}}{M} F_{\alpha 2} + \frac{M_{13}}{M} F_{\alpha 3}.
\end{equation}
The second and third terms on the right hand side of \eqref{eq:F1tilde} are contributions to $\tilde{F}_{\alpha 1}$ induced by the mixing of $N_1$ with $N_2$ and $N_3$ to form the mass eigenstate $N_1^m$. Direct searches for the X/$\gamma$-ray line corresponding to this decay provide the strongest limits on $\theta_1$ (as a function of $M_1$) for the mass range relevant to the $\nu$MSM. A summary of these limits is given in~\cite{Wat12}. In general,
\begin{equation}
\label{eq:xray}
\theta_1^2 \lesssim 3 \times 10^{-5} \paren{\frac{\text{keV}}{M_1}}^5
\end{equation}
must be satisfied for $0.5~\text{keV} \lesssim M_1 \lesssim 14~\text{MeV}$, though the constraint is typically 100 times stronger than \eqref{eq:xray} for masses outside the 12--40~keV range~\cite{Wat12}. For $N_1^m$ produced entirely from active-sterile neutrino mixing, $M_1$ can be bounded above by combining the X-ray constraints with the requirement of sufficient dark matter production ($\propto \theta_1^2$). The bound obtained depends on the lepton asymmetry at the time of $N_1^m$ production: a negligible lepton asymmetry is called the non-resonant production (NRP) scenario while a large lepton asymmetry is called the resonant production (RP) scenario. The bounds for these two scenarios are~\cite{Sha07a,Wat12,Sha08}
\begin{equation}
\label{eq:M1}
M_1^\text{NRP} \lesssim 2.2\text{~keV}, \quad M_1^\text{RP} \lesssim 40\text{~keV}.
\end{equation}
Meanwhile, $M_1$ can be bounded below by phase-space density arguments for dwarf spheroidal galaxies~\cite{Boy09b,Gor10}, the Lyman-$\alpha$ forest data~\cite{Boy09c,Boy09d}, studies of gravitationally lensed QSOs~\cite{Mir07}, and N-body simulations of the Milky Way~\cite{Pol11}. The bounds from the Lyman-$\alpha$ forest data and N-body simulations of the Milky Way are the strongest and give\footnote{These are the (Bayesian) $2 \sigma$ bounds. Although \cite{Pol11} quotes a stronger bound for $M_1^\text{RP}$, it is based on a simple mass rescaling argument that is shown to be insufficient for a more rigorous analysis of the Lyman-$\alpha$ forest bound in the RP scenario~\cite{Boy09d}.}
\begin{equation}
\label{eq:lymanalpha}
M_1^\text{NRP} \gtrsim 13\text{~keV}, \quad M_1^\text{RP} \gtrsim 2\text{~keV}.
\end{equation}
Combining \eqref{eq:M1} and \eqref{eq:lymanalpha} rules out the simpler NRP scenario, even with a possibly large entropy dilution from the decays of $N_2^m$ and $N_3^m$~\cite{Sha07a}. The RP scenario is still allowed for a range of $M_1$; it requires an even larger degeneracy than \eqref{eq:deltaM}, on the order $\Delta M \lesssim 10^{-7}$~eV, to produce the required lepton asymmetry for enhanced dark matter production~\cite{Roy10}. This level of degeneracy is unstable in the presence of radiative corrections and must be achieved with either fine-tuning or an extension of the model by a Planck-scale symmetry and non-renormalizable operators~\cite{Roy10}.

\subsection{Dark matter production from a Higgs singlet}

An alternative dark matter production scenario that is capable of satisfying the Lyman-$\alpha$ forest bound for warm dark matter (or allows for heavier cold dark matter) uses a real Higgs singlet $\sgt$ and its decays to $N_1^m$~\cite{Sha06}. This scenario is arguably simpler than the RP scenario and has the advantage that Majorana masses originate from the VEV of $\sgt$ rather than as bare mass terms. This extension of the $\nu$MSM, which we will call the \emph{neutrino Next-to-Minimal Standard Model} ($\nu$NMSM), is the basis of this paper.

In the $\nu$NMSM, the Majorana mass term in the Lagrangian~\eqref{eq:lagrangian} is modified to
\begin{equation}
\label{eq:lambda}
\Delta \mathcal{L} = -\frac{\lambda_{IJ}}{2}\sgt \bar N_I^c N_J,
\end{equation}
where $M_{IJ} = \lambda_{IJ} \left< \sgt \right>$ once $\sgt$ acquires a VEV. In the mass basis $N_I^m$, $\lambda_{IJ} \rightarrow \lambda_I$ where $M_I = \lambda_I \left<\sgt\right>$. The mixing angle $\theta_1^2$ is assumed small enough that dark matter production from active-sterile neutrino mixing is negligible and the X/$\gamma$-ray constraint \eqref{eq:xray} is satisfied.\footnote{Since the NRP and RP bounds \eqref{eq:M1} no longer apply, $M_1$ may exceed the range in which \eqref{eq:xray} is valid. In this case, $\gamma$-ray constraints from EGRET~\cite{Ber07} and FERMI~\cite{Cir12} give $\tau \gtrsim 10^{26}$~s, or equivalently $\theta_1^2 \lesssim 2 \times 10^{-20} \paren{\text{MeV}/{M_1}}^5$, for masses up to 30~TeV.} 
Assuming no miraculous cancellations of terms in \eqref{eq:F1tilde}, this requires
\begin{equation}
\label{eq:xray3}
F_{\alpha 1}, \frac{M_{12}}{M}F_{\alpha 2}, \frac{M_{13}}{M}F_{\alpha 3} \lesssim 10^{-13}.
\end{equation}
With \eqref{eq:xray3}, one can show that the induced contributions to $M_1$ from $M_{12}$ and $M_{13}$ are small~\cite{Sha07} and hence $\lambda_1 \simeq \lambda_{11}$.
Dark matter production then proceeds via the decays $\sgt^m \rightarrow N_1^m N_1^m$ with the partial width~\cite{Sha06}
\begin{equation}
\label{eq:gamma}
\Gamma = \frac{\lambda_{1}^2}{16\pi}m_\sgt \simeq \frac{\lambda_{11}^2}{16\pi}m_\sgt,
\end{equation}
where $m_\sgt > 2 M_1$ is the mass of the physical mass eigenstate $\sgt^m$.\footnote{We have assumed a small mixing angle $\theta_{h \sgt}$ between the Higgs boson $h$ and $\sgt$ so that $\sgt^m \simeq \sgt$, $h^m \simeq h$, and the decays $h^m \rightarrow N_1^m N_1^m$ are negligible compared to $\sgt^m \rightarrow N_1^m N_1^m$~\cite{Kus08}. This is a good approximation for both models considered in section~\ref{sec:sym}.}
This production depends on the thermal history of $\sgt^m$, specifically the ratio of its mass to its freeze-out temperature, $r_f \equiv m_\sgt / T_f$~\cite{Kus08}. For the case that $\sgt^m$ is in thermal equilibrium down to temperatures $T \ll m_\sgt$ (i.e.\ $r_f \gg 1$), the dark matter abundance is given by~\cite{Sha06}
\begin{equation}
\label{eq:omegas}
\Omega_{N_1^m} \simeq \frac{0.2 f(m_\sgt)}{S} \paren{\frac{\lambda_{11}}{10^{-10}}}^2 \paren{\frac{ M_1}{4\text{~keV}}} \paren{\frac{\text{GeV}}{m_\sgt}},
\end{equation}
where $f(m_\sgt) \simeq \paren{10.75/g_*\paren{m_\sgt / 3}}^{3/2}$ and $1 \leq S \lesssim 2$ is a factor that accounts for entropy production from the decays of $N_2^m$ and $N_3^m$ after $N_1^m$ is produced.\footnote{Since $N_1^m$ production peaks at $T_\text{prod} \equiv m_\sgt / 2.3$, \eqref{eq:omegas} is a good approximation for $r_f \gtrsim 3$~\cite{Kus08}.}
Using $M_1 \simeq \lambda_{11} \left< \sgt \right>$ in \eqref{eq:omegas}, the appropriate dark matter abundance $\Omega_{N_1^m} \simeq 0.23$ is generated when
\begin{equation}
\label{eq:lambda11}
\lambda_{11} \simeq 4 \times 10^{-9} \paren{\frac{ S}{f(m_\sgt)}}^{1/3} \left( \frac{m_\sgt}{\left< \sgt \right>} \right) ^ {1/3}.
\end{equation}
For the case that $\sgt^m$ is a thermal relic decaying out of equilibrium (i.e.\ $r_f \ll 1$), the dark matter abundance is given by~\cite{Kus08}
\begin{equation}
\label{eq:omegaN1ooe}
\Omega_{N_1^m} \simeq \frac{0.3}{S} \paren{\frac{M_1}{\text{keV}}} \paren{\frac{10.75}{g_*(T_f)}} \paren{\frac{B}{0.01}},
\end{equation}
where $B \equiv \Gamma / \Gamma_\sgt^\text{tot}$ is the branching ratio of $\sgt^m \rightarrow N_1^m N_1^m$.\footnote{As in~\cite{Kus08}, we neglect any $\sgt \sgt$ annihilations that could reduce \eqref{eq:omegaN1ooe} by up to a factor of 2.} Analytic expressions relevant to the intermediate case $r_f \sim 1$ can be found in~\cite{Kus08}, and the result is a combination of \eqref{eq:omegas} and \eqref{eq:omegaN1ooe}.

The Lyman-$\alpha$ forest bound for this dark matter production mechanism can be estimated by rescaling the NRP bound, giving~\cite{Sha07a,Kus08}
\begin{equation}
\label{eq:lymanalphahiggs}
M_1^\text{Higgs} \gtrsim 10 \paren{\frac{10.75}{g_*(T_\text{prod})}}^{1/3}\text{~keV},
\end{equation}
where $T_\text{prod}$ is the temperature at which $N_1^m$ is produced. Further constraints come from the requirement that the interactions $\sgt^m \leftrightarrow N_2^m N_2^m$ and $\sgt^m \leftrightarrow N_3^m N_3^m$ (and any interactions $\text{SM} \leftrightarrow N_2^m N_2^m$ and $\text{SM} \leftrightarrow N_3^m N_3^m$ mediated by $\sgt^m$) do not bring $N_2^m$ and $N_3^m$ into thermal equilibrium at the characteristic temperature of leptogenesis~\cite{Asa12}
\begin{equation}
T_L \sim \paren{ \frac{M \Delta M M_0}{3} }^{1/3},
\end{equation}
where $M_0 \simeq 7 \times 10^{17}$~GeV, and spoil baryogenesis.\footnote{If these interactions bring $N_2^m$ and $N_3^m$ into thermal equilibrium below $T_L$, the asymmetry in the sterile neutrinos will be wiped out but the asymmetry in the active neutrinos will remain.}
Moreover, the addition of $\sgt$ must not open an invisible branching ratio of the Higgs greater than 30\% at 2$\sigma$~\cite{Gia12}.
These constraints are discussed further in section~\ref{sec:sym} for specific models of the scalar sector.

Although we have assumed that $\sgt$ is real in the discussion above, it is also possible (with some restrictions) to have a complex $\sgt$. (We parametrize $\sgt = \paren{\rho + i\GB} / \sqrt{2}$ for a complex $\sgt$ but continue to use $m_\sgt$ and $\sgt^m$ instead of $m_\rho$ and $\rho^m$ to maintain consistency with the notation for a real $\sgt$.) In previous studies of the $\nu$NMSM, which do not attempt to explain the origin of its parameters, $\sgt$ is typically assumed real to avoid a massless Goldstone boson~$\GB$ and hence the unsuitably fast decay channel $N_1^m \rightarrow \nu^m \GB$ for dark matter~\cite{Sha06,Ani09,Bez10,Kus06,Kus08}. We have found it very difficult, however, to explain the parameters of the $\nu$NMSM with an underlying symmetry if $\sgt$ is real and hence uncharged. To construct such a symmetry, we must therefore consider a complex $\sgt$ and address the problems and constraints associated with a Goldstone boson.

There are several ways to avoid the decay $N_1^m \rightarrow \nu^m \GB$ for a complex $\sgt$. First, if $\sgt$ is charged under a discrete symmetry and terms of the form $\sgt^n + \sgt^{\dag n}$ are allowed, these terms give $\GB$ a mass and can kinematically forbid the decay $N_1^m \rightarrow \nu^m \GB$. If the analogous decays $N_2^m,N_3^m \rightarrow \nu^m \GB$ are still allowed, they can relax the constraint~\eqref{eq:M} to $M \gtrsim \text{few}$~MeV~\cite{Sha07}. Alternatively, if $\GB$ remains light enough to allow $N_1^m \rightarrow \nu^m \GB$ then the mixing of $N_1$ with other neutrino species can be suppressed or forbidden by a symmetry, thereby suppressing the decay.
This case is particularly interesting since $\GB$ can contribute to the effective number of neutrino species and give a value of $N_\text{eff}$ above the SM or $\nu$MSM prediction, as recent measurements prefer (see~\cite{Aba12} and references therein).\footnote{The real component of $\sgt$ can also contribute to $N_\text{eff}$ during BBN if $m_\sgt \lesssim 10$~MeV~\cite{Boe12}. For the models of the scalar sector considered in section~\ref{sec:sym}, however, $m_\sgt \gg 10$~MeV.} The contribution of $\GB$ to $N_\text{eff}$ depends on the freeze-out temperature $T_f$: it can be as large as $\Delta N_\text{eff} \sim 1$ for a thermal distribution of $\GB$ or much smaller if $\GB$ decouples early. The Planck experiment and other future cosmic microwave background (CMB) experiments will therefore be able to constrain these models with a complex $\sgt$~\cite{Gal10}.

\section{Symmetries and the \boldmath{$\nu$}NMSM}
\label{sec:sym}
The $\nu$NMSM, like the $\nu$MSM, requires parameters that are constrained to be hierarchically small. An important question is whether it is possible for such structure to come from an underlying symmetry. In the context of the $\nu$MSM, flavour symmetries~\cite{Sha07,Mer11b,Ara12}, the split seesaw mechanism~\cite{Kus10,Adu11}, and the Froggatt-Nielsen mechanism~\cite{Mer11,Bar11,Bar12} have been explored for producing the required pattern of Majorana masses and Yukawa couplings. Following this approach, we would like to demonstrate explicitly how the parameters of the $\nu$NMSM can arise from symmetries broken at or near the Planck scale.
Since the values of some parameters (e.g.\ $\lambda_{11}$ in \eqref{eq:lambda11}) depend on an unspecified scalar sector, we first keep the discussion general and then consider two specific models of the scalar sector: one in which $\sgt$ helps stabilize the electroweak vacuum~\cite{Eli12} and one in which $\sgt$ is the inflaton~\cite{Sha06,Ani09,Bez10}. These models of the scalar sector, though motivated as minimal solutions to other possible problems with the $\nu$MSM, are meant only to provide definite examples for the symmetries used in the flavour sector; other models may certainly be considered. We do not provide an explanation for the values of parameters in the scalar sector or the associated hierarchy problems since little is known about their origin.

\subsection{Symmetries in the flavour sector}
First consider how the structure of the $\nu$NMSM Lagrangian,
\begin{equation}
\label{eq:lagrangian2}
\Delta \mathcal{L} = - {F_{\alpha I}}{\bar L_\alpha }{N_I}\Higgs - \frac{{{\lambda_{IJ}}}}{2}\sgt\bar N_I^c{N_J} + {\text{h.c.}},
\end{equation}
can arise from an underlying symmetry without regard to the size of the couplings $\FaI$ and $\lambda_{IJ}$.
There are several ways this structure can arise:
\begin{itemize}
\item \emph{Conformal symmetry/scale invariance}: The structure \eqref{eq:lagrangian2}, which has only terms with dimensionless couplings, can arise from models with a classical conformal symmetry~\cite{Nic07,Foo07,Nic08} or hidden scale invariance~\cite{Buc87,Buc91}. These models have been motivated as a solution to the hierarchy problem: the conformal symmetry forbids tree-level scalar mass terms while radiative breaking of this symmetry by the conformal anomaly is responsible for electroweak symmetry breaking and, in \cite{Buc87,Buc91}, a hierarchy between the electroweak and Planck scales from a choice of large scale $f$. Unfortunately, existing models of this type are not fully realistic.
\item \emph{(Approximate) Global U(1) symmetry}: For a complex $\sgt$, the structure \eqref{eq:lagrangian2} can arise from a global U(1) symmetry under which $\sgt$ is charged. (We use a global symmetry to avoid introducing a new low-energy gauge sector.) Since it has been argued that the only symmetries allowed in an effective low-energy theory are those that derive from gauge symmetries~\cite{Iba91}, note that approximate global symmetries (approximate because they are broken by non-perturbative effects) can arise from string theory as the remnant of a non-linearly realized U(1) gauge symmetry in which the gauge boson acquires a large (string scale) mass through its coupling to a Stueckelberg field~\cite{Bur08}. For a consistent model, the underlying U(1) gauge symmetry must be anomaly-free or Green-Schwarz anomalous~\cite{Iba01,Ant02}. An anomaly-free example in which matter fields have U(1)$_{B-L}$ charges
is given in table~\ref{table:symmetries}.
\item \emph{Discrete $Z_N$ symmetry}: A discrete $Z_N$ symmetry can also give the structure \eqref{eq:lagrangian2}. Such symmetries can arise from the spontaneous breaking of a gauge symmetry at a high scale~\cite{Kra89} or from coupling selection rules on heterotic orbifolds (see~\cite{Ara08} and references therein). Note that it is often easier to satisfy the anomaly cancellation conditions for $Z_N$ symmetries~\cite{Ara07,Ara08} than those for U(1) symmetries: an anomaly-free $Z_3$ example is given in table~\ref{table:symmetries}.\footnote{The mixed $Z_N$-U(1)$_Y$-U(1)$_Y$ anomaly does not pose a meaningful constraint since the hypercharge normalization is not fixed~\cite{Iba93,Dre06}.} However, the spontaneous breaking of discrete symmetries can produce domain walls~\cite{Zel74} and care must be taken to avoid these,
such as by having the symmetry breaking phase transition occur below 1~MeV~\cite{Cas01}.
\end{itemize}
\begin{table}[t]
\centering
\begin{tabular}{|c|ccccc|ccc|cc|}
\hline
   & $N_1$ & $N_2$ & $N_3$ & $L_\alpha$ & $E_\alpha$ & $Q_i$ & $U_i$ & $D_i$ & $\Higgs$ & $\sgt$ \\
  \hline
  U(1) & -1 & -1 & -1 & -1 & -1 & 1/3 & 1/3 & 1/3 & 0 & 2 \\
  $Z_3$ & 1 & 1 & 1 & 1 & 1 & 0 & 0 & 0 & 0 & 1 \\
  \hline
\end{tabular}
\caption{\label{table:symmetries} Examples of an anomaly-free global U(1) and $Z_3$ symmetry that can give the Lagrangian structure \eqref{eq:lagrangian2}. Note: $E_\alpha$ are the right-handed charged leptons, $Q_i$ ($i=1,2,3$) are the left-handed quark doublets, and $U_i, D_i$ are the right-handed quarks.}
\end{table}
Although either a global U(1) or discrete $Z_N$ symmetry can give the desired Lagrangian structure~\eqref{eq:lagrangian2}, we use a global U(1) symmetry to avoid introducing the problems associated with domain walls.

Now consider the hierarchy of Majorana masses and Yukawa couplings in the $\nu$NMSM. To explain the small Yukawa couplings $\tilde{F}_{\alpha 1} \lesssim 10^{-13}$ and, for a complex $\sgt$, to prevent the fast dark matter decay channel $N_1^m \rightarrow \nu^m \GB$,
we introduce a $Z_2$ symmetry under which only $N_1$ is charged (see table~\ref{table:FN1}).\footnote{The anomaly cancellation conditions for this $Z_2$ are trivially satisfied. Therefore this symmetry is exact at the quantum level.} This symmetry allows only the couplings
\begin{equation}
\label{eq:matrices}
F_{\alpha I} = \left(
\begin{matrix}
  0&F_{e 2}&F_{e 3}\\
  0&F_{\mu 2}&F_{\mu 3}\\
  0&F_{\tau 2}&F_{\tau 3}
\end{matrix}
\right), \quad
\lambda_{IJ} = \left(
\begin{matrix}
  \lambda_{11} & 0 & 0 \\ 
  0 & \lambda_{22} & \lambda_{23} \\ 
  0 & \lambda_{23} & \lambda_{33} \\
\end{matrix}
\right),
\end{equation}
and hence forbids mixing of $N_1$ with the other neutrinos, making $N_1^m$ completely stable ($\theta_1 = 0$) and one active neutrino exactly massless. The required pattern of Majorana masses and Yukawa couplings can then be produced if there are strong hierarchies in the remaining $\lambda_{IJ}$ and $\FaI$, specifically if
\begin{equation}
\label{eq:couplings}
\begin{gathered}
F_{\alpha 2} \sim F_2, \quad F_{\alpha 3} \sim F_3,\\
\lambda_{11} \sim \frac{M_1}{\left<\sgt\right>}, \quad \lambda_{23} \sim \frac{M}{\left<\sgt\right>}, \quad \max \left\{ \lambda_{22}, \lambda_{33}\right\} \sim \frac{\Delta M}{\left<\sgt\right>}.
\end{gathered}
\end{equation}
We consider two possibilities for generating these hierarchies from an underlying symmetry, in which case the small couplings in \eqref{eq:couplings} are preserved under the renormalization group flow:
\begin{itemize}
\item \emph{Froggatt-Nielsen mechanism}: The Froggatt-Nielsen mechanism~\cite{Fro79} is a well-known method of generating hierarchical parameters. In brief, a new U(1)$_\text{FN}$ gauge symmetry that is spontaneously broken by a flavon field $\vartheta$ at a very high scale is introduced. Fields of the $\nu$NMSM are charged under this U(1)$_\text{FN}$ so that $\vartheta$ (or $\vartheta^\dag$) must couple to the terms in \eqref{eq:lagrangian2} with various powers to form gauge singlets. After the U(1)$_\text{FN}$ is spontaneously broken, these non-renormalizable terms are suppressed by powers of $\eta \equiv \left< \vartheta \right>/\Mp$, where $\eta$ is a free parameter (though typically assumed to be on the order of the Cabibbo angle~\cite{Mer11,Dat05}). Of course, multiple flavon fields $\vartheta_i$ with various $\eta_i \equiv \left< \vartheta_i \right> / \Mp$  may be used, as well as a discrete $Z_N$ symmetry in place of the U(1)$_\text{FN}$.
\item \emph{Non-perturbative symmetry breaking}: Another possibility for generating hierarchical parameters comes from non-perturbative symmetry breaking in string theory. In~\cite{And12}, for example, it is shown that heterotic string compactifications on Calabi-Yau manifolds can give models with the SM gauge group and additional U(1) symmetries. These additional symmetries can play a role analogous to that of the U(1)$_\text{FN}$: if the $\nu$NMSM fields are charged under these symmetries, the terms in \eqref{eq:lagrangian2} may require couplings to various powers of $\vartheta_i \equiv e^{-T^i/\Mp}$ to form gauge singlets, where $T^i = t^i + 2i \chi^i$ are K{\"a}hler moduli with axionic components $\chi^i$ (not to be confused with the Goldstone boson $\GB$) that transform non-linearly under the U(1). After these symmetries are spontaneously broken by $\left< t^i \right> \gg \Mp$ \cite{And11,And12}, the terms in \eqref{eq:lagrangian2} are suppressed by powers of $\eta_i \equiv e^{-\left< t^i \right>/\Mp}$. Again, discrete $Z_N$ symmetries may be used in place of the U(1) symmetries.
\end{itemize}
Although either mechanism may be used to generate the hierarchical parameters \eqref{eq:couplings} for the same charge assignment, the non-perturbative symmetry breaking mechanism does not require additional symmetry breaking or scalar particles below the Planck scale and therefore adheres closer to the ``minimal'' philosophy of the $\nu$MSM.

To fix the absolute scale of the couplings $\lambda_{IJ}$ and hence construct an explicit model of symmetries in the flavour sector, the values of $m_\sgt$ and $\left<\sgt\right>$ must be fixed (see \eqref{eq:lambda11} and \eqref{eq:couplings}) by some model of the scalar sector. We now consider two models of the scalar sector that are motivated as solutions to other possible problems with the $\nu$MSM.

\subsection{Stabilization of the electroweak vacuum}
For a Higgs mass $m_h \simeq 125$--126~GeV, the SM (and hence $\nu$MSM) potential develops an instability below the Planck scale unless the top mass is about 2$\sigma$ below its central value~\cite{Eli11}. While it is possible that more precise measurements of the top mass will lower its central value and relieve this tension, we first consider a model of the scalar sector in which the Higgs singlet can, for the central value of the top mass, stabilize the electroweak vacuum through a scalar threshold effect.

This model, described in~\cite{Eli12}, considers a complex $\sgt$ and scalar potential of the form
\begin{equation}
\label{eq:Vthreshold}
V = \lambda_h \paren{\Higgs^\dag \Higgs - \frac{v^2}{2}}^2 + \lambda_\sgt \paren{\sgt^\dag \sgt - \frac{w^2}{2}}^2 + 2\lambda_{h \sgt}\paren{\Higgs^\dag \Higgs - \frac{v^2}{2}}\paren{\sgt^\dag \sgt - \frac{w^2}{2}},
\end{equation}
which is the most general renormalizable potential that respects a global abelian symmetry under which only $\sgt$ is charged. Values of $\lambda_h, \lambda_\sgt > 0$ and $\lambda_{h \sgt}^2 < \lambda_h \lambda_\sgt$ are assumed so that the minimum of this potential is given by
\begin{equation}
\label{eq:Vmin}
\left< \Higgs^\dag \Higgs \right> = \frac{v^2}{2}, \quad \left< \sgt^\dag \sgt \right> = \frac{w^2}{2},
\end{equation}
where $v = 246$~GeV. The mass matrix for the real components of $\Higgs$ and $\sgt$ is then
\begin{equation}
\label{eq:twoeig}
\mathcal{M}^2 = 2 \paren{\begin{matrix}
\lambda_h v^2 & \lambda_{h \sgt} v w\\
\lambda_{h \sgt} vw & \lambda_\sgt w^2\\
\end{matrix}},
\end{equation}
while the imaginary component of $\sgt$ (i.e.\ $\GB$) remains massless. In contrast to other models that use a Higgs singlet to stabilize the electroweak vacuum (e.g.\ \cite{Che12,Leb12}), this model assumes $w \gg v$. The two eigenstates of \eqref{eq:twoeig} then have masses
\begin{align}
m_h^2 &= 2v^2 \left[ \lambda_h - \frac{\lambda_{h \sgt}^2}{\lambda_\sgt} + \mathcal{O}\paren{\frac{v^2}{w^2}}\right],\label{eq:ratio1} \\
m_\sgt^2 &= 2w^2 \left[ \lambda_\sgt + \frac{\lambda_{h \sgt}^2}{\lambda_\sgt}\paren{\frac{v^2}{w^2}} + \mathcal{O}\paren{\frac{v^4}{w^4}} \right],\label{eq:ratio2}
\end{align}
with a mixing angle $\theta_{h \sgt} \sim v/w$.
Integrating out the heavier state for scales below $m_\sgt$ gives the effective potential
\begin{equation}
V_\text{eff} = \lambda \paren{\Higgs^\dag \Higgs - \frac{v^2}{2}}^2, \quad \lambda \equiv \lambda_h - \frac{\lambda_{h \sgt}^2}{\lambda_\sgt},
\end{equation}
where the matching condition for the Higgs quartic coupling gives a tree-level shift $\delta \lambda \equiv \lambda_{h \sgt}^2/\lambda_\sgt$ from $\lambda$ just below $m_\sgt$ to $\lambda_h$ just above $m_\sgt$. Provided $m_\sgt$ is below the instability scale $\Lambda \simeq 10^9$--$10^{10}$~GeV~\cite{Eli11}, a value of $\delta \lambda \simeq 0.01$ can push the instability beyond the Planck scale.

Due to the massless Goldstone boson $\GB$, the value of $\lambda_{h \sgt}$ is constrained by limits on the invisible branching ratio of the Higgs. For $m_h \simeq 125$~GeV, the total SM decay width of the Higgs is~\cite{Den11}
\begin{equation}
\Gamma_\text{SM} = 4.07\text{~MeV},
\end{equation}
while the invisible decay width for $h^m \rightarrow \GB \GB$ is~\cite{Leb11}
\begin{equation}
\Gamma_\text{inv} = \frac{\lambda_{h \sgt}^2 v^2}{8\pi m_h}.
\end{equation}
Allowing an invisible branching ratio of up to 30\%~\cite{Gia12} gives the constraint
\begin{equation}
\label{eq:yhs}
\lambda_{h\sgt}(m_h) \lesssim 0.01.
\end{equation}
A value of $\delta \lambda$ that stabilizes the electroweak vacuum and is consistent \eqref{eq:yhs} can then be obtained for $\lambda_\sgt \lesssim 0.01$ (the running of $\lambda_{h\sgt}$ and $\lambda_\sgt$ is small for these values). We illustrate this by constructing a model with $\lambda_{h \sgt}, \lambda_\sgt \sim 0.01$ and hence an invisible branching ratio of the Higgs of about 30\%.\footnote{It is, however, quite simple to construct models with a smaller invisible branching ratio by taking smaller $\lambda_{h\sgt}$ and $\lambda_\sgt$ while keeping $\delta \lambda$ fixed.} For these values, one can show that $\GB$ remains in thermal equilibrium down to temperatures just below $m_\mu$. The model therefore has a $\Delta N_\text{eff} \simeq 4/7$ contribution to the effective number of neutrino species from $\GB$ and hence a total value of $N_\text{eff} \simeq 3.6$. This value can be tested by the Planck experiment and other future CMB experiments~\cite{Gal10}.

Now consider the flavour sector of the $\nu$NMSM for this model of the scalar sector. For $\lambda_{h \sgt} \sim 0.01$, the interactions $\Higgs^\dag \Higgs \leftrightarrow \sgt^m \sgt^m$ keep $\sgt^m$  (the real component of $\sgt$) in thermal equilibrium down to temperatures $T \ll m_\sgt$ for any mass $m_\sgt \lesssim \Lambda$. We are therefore in the dark matter production case $r_f \gg 1$. 
For $\lambda_\sgt \sim 0.01$, the ratio $m_\sgt / \left<\sgt\right>$ is fixed by \eqref{eq:ratio2} and \eqref{eq:lambda11} gives a value of
\begin{equation}
\label{eq:lambda11stab}
\lambda_{11} \sim 1 \times 10^{-8}
\end{equation}
to produce the correct dark matter abundance.\footnote{Here we have used $S \simeq 1$ (anticipating $M \sim 1$~GeV) and taken $m_\sgt \gtrsim \TEW$ for $f\paren{m_\sgt}$. Also, $\lambda_{11}$ must be a factor of $\sqrt{2}$ larger than in \eqref{eq:lambda11} for a complex $\sgt$ since only the real component of $\sgt$ can decay to $N_1^m$.} The Lyman-$\alpha$ forest bound \eqref{eq:lymanalphahiggs} is therefore satisfied for a choice $\left<\sgt\right> \gtrsim 500$~GeV. 
Taking $\left< \sgt \right> \simeq 10^8$~GeV, a pattern of masses $M_{IJ}$ and couplings $\FaI$ that gives the correct dark matter abundance and baryon asymmetry can be achieved with two fields $\vartheta_1$, $\vartheta_2$, the values $\eta_1 \simeq 10^{-8}$, $\eta_2 \simeq 10^{-7}$, and the charge assignments given in table~\ref{table:FN1}.
\begin{table}[t]
\centering
\begin{tabular}{|c|ccccc|ccc|cc|cc|}
  \hline
  & $N_1$ & $N_2$ & $N_3$ & $L_\alpha$ & $E_\alpha$ & $Q_i$ & $U_i$ & $D_i$ & $\Higgs$ & $\sgt$ & $\vartheta_1$ & $\vartheta_2$ \\
  \hline
  U(1) & -1 & -1 & -1 & -1 & -1 & 1/3 & 1/3 & 1/3 & 0 & -1 & 3 & 0 \\
  $Z_3$ & 0 & 1 & -1 & 0 & 0 & 0 & 0 & 0 & 0 & 0 & 0 & 1 \\
  $Z_2$ & 1 & 0 & 0 & 0 & 0 & 0 & 0 & 0 & 0 & 0 & 0 & 0 \\
  \hline
\end{tabular}
\caption{\label{table:FN1} Charge assignments for the stabilization of the electroweak vacuum scenario.
The global U(1) symmetry gives the structure \eqref{eq:lagrangian2} while the discrete $Z_3$ and $Z_2$ symmetries, together with the fields $\vartheta_1$ and $\vartheta_2$, give the required hierarchies in $F_{\alpha I}$ and $\lambda_{IJ}$.
}
\end{table}
We stress that this is the simplest anomaly-free model we could find, though other charge assignments are possible.\footnote{The $Z_3$ and $Z_2$ symmetries could be combined in a single $Z_6$, if desired.}

For the sake of definiteness, suppose that the non-perturbative symmetry breaking mechanism is used for generating the hierarchies in $F_{\alpha I}$ and $\lambda_{IJ}$; that is, $\vartheta_i = e^{-T^i/\Mp}$ and $\eta_i = e^{-\left<t^i\right>/\Mp}$ for $i = \left\{1,2\right\}$. From table~\ref{table:FN1}, the Lagrangian for the flavour sector is then
\begin{equation}
\label{eq:lagrangianVacuumStability}
\begin{aligned}
\Delta \mathcal{L} &= - f_{\alpha 2} \vartheta_2^\dag {\bar L_\alpha }{N_2}\Higgs - f_{\alpha 3} \vartheta_2 {\bar L_\alpha }{N_3}\Higgs - \frac{h_{11}}{2} \vartheta_1 \sgt \bar N_1^c N_1 - \frac{h_{22}}{2} \vartheta_1 \vartheta_2 \sgt \bar N_2^c N_2 \\
& \qquad - \frac{h_{23}}{2} \vartheta_1 \sgt \bar N_2^c N_3 - \frac{h_{32}}{2} \vartheta_1 \sgt \bar N_3^c N_2 - \frac{h_{33}}{2} \vartheta_1 \vartheta_2^\dag \sgt \bar N_3^c N_3 + \text{h.c.},
\end{aligned}
\end{equation}
where $f_{\alpha I}$ and $h_{IJ}$ are $\mathcal{O}(1)$ couplings. Meanwhile, the scalar potential is given by \eqref{eq:Vthreshold} plus the additional terms
$\vartheta_i^\dag \vartheta_i H^\dag H$, $\vartheta_i^\dag \vartheta_i \sgt^\dag \sgt$, and $\vartheta_1 \sgt^3 + \vartheta_1^\dag \sgt^{\dag3}$ involving $\vartheta_1$ and $\vartheta_2$. Note that we must assume these additional terms, which are allowed by the symmetries, have sufficiently small coefficients to preserve \eqref{eq:Vthreshold}.\footnote{The $\vartheta_1 \sgt^3 + \vartheta_1^\dag (\sgt^\dag)^3$ terms, in particular, give $\GB$ a small mass and lead to the formation of a discrete $Z_3$ symmetry in $\sgt$ after the spontaneous symmetry breaking associated with $\vartheta_1$, which can introduce potentially dangerous domain walls when this $Z_3$ is later broken by the VEV of $\sgt$~\cite{Cas01}.} For the former two terms, this assumption is a further aspect of the hierarchy problem in the scalar sector. An explanation for these small coefficients may arise from the solution to the hierarchy problem, but providing such an explanation goes beyond the scope of this paper. It is interesting to see, however, that in order to produce hierarchical parameters in the flavour sector of the $\nu$NMSM with symmetries the hierarchy problem in the scalar sector may be made worse.\footnote{This point is particularly relevant to the recent work~\cite{Lyn13}, which has suggested that the Higgs mass in the SM does not have the quadratic divergence that is usually identified with the hierarchy problem. In this case, trying to explain the hierarchical parameters of any model with additional high-energy scalars may reintroduce the hierarchy problem.}
For the latter terms $\vartheta_1 \sgt^3 + \vartheta_1^\dag \sgt^{\dag3}$, we similarly accept a small parameter in the scalar sector without explanation, but note that these terms could also be forbidden by an additional U(1) symmetry under which $\sgt$ and $\vartheta_1$ have opposite charges.

After the spontaneous symmetry breaking associated with $\vartheta_1$ and $\vartheta_2$,
\eqref{eq:lagrangianVacuumStability} reduces to \eqref{eq:lagrangian2} with the textures
\begin{equation}
\label{eq:texture1}
{F_{\alpha I}} \sim \left(
\begin{matrix}
  0 & \eta_2 & \eta_2 \\
  0 & \eta_2 & \eta_2 \\
  0 & \eta_2 & \eta_2
\end{matrix}
\right), \quad
{\lambda_{IJ}} \sim \left(
\begin{matrix}
  \eta_1 & 0 & 0 \\ 
  0 & \eta_1 \eta_2 & \eta_1 \\ 
  0 & \eta_1 & \eta_1 \eta_2 \\
\end{matrix}
\right).
\end{equation}
The parameters of the $\nu$NMSM are then
\begin{equation}
\label{eq:param1}
\begin{gathered}
F_{\alpha 2} \sim 1 \times 10^{-7}, \quad F_{\alpha 3} \sim 1 \times 10^{-7},\\
M_1 \sim 1\text{~GeV}, \quad M \sim 1\text{~GeV}, \quad \Delta M \sim 100\text{~eV},
\end{gathered}
\end{equation}
up to $\mathcal{O}\paren{1}$ constants. This example shows that, in contrast to the $\nu$MSM, dark matter in the $\nu$NMSM can be much heavier than the keV scale.
Active neutrino mixing in this model is anarchical (up to charged lepton corrections) while the charged lepton and quark Yukawa couplings remain unsuppressed. Therefore additional flavour symmetries using the Green-Schwarz anomaly cancellation mechanism, such as in \cite{Kan05,Mer11}, must be used to produce hierarchies in the charged lepton and quark sectors.

As a consistency check on this model, we must verify that $N_2^m$ and $N_3^m$ are out of thermal equilibrium at the characteristic temperature of leptogenesis $T_L \sim 3 \times 10^3$~GeV. Since $m_\sgt \simeq 2 \times 10^7\text{~GeV} \gg T_L$, $\sgt^m$ has decayed away by leptogenesis\footnote{Note that a relic CP-even distribution of $N_2^m$ and $N_3^m$ from the decays of $\sgt^m$ does not affect leptogenesis.} and only the scattering processes $\Higgs^\dag \Higgs \leftrightarrow N_2^m N_2^m$ and $\Higgs^\dag \Higgs \leftrightarrow N_3^m N_3^m$ mediated by $\sgt^m$ and $\GB$ need to be considered. For $\lambda_{h \sgt} \sim 0.01$, these processes are out of equilibrium at $T_L$ for $\lambda_2, \lambda_3 \simeq \lambda_{23} \lesssim 10^{-5}$, which is satisfied by \eqref{eq:texture1}.

This model demonstrates that it is possible to use symmetries broken at or near the Planck scale to obtain the hierarchical pattern of Majorana masses and Yukawa couplings required for successful baryogenesis and dark matter production in the $\nu$NMSM. The model obeys all phenomenological constraints and 
allows for the possibility of Higgs inflation by ensuring that the Higgs potential does not develop a second minimum before the Planck scale. Of course, the symmetries used do not address the hierarchy problem associated with radiative corrections to the scalar sector. To do so would involve implementing a supersymmetric version of the theory, which departs from the underlying philosophy of the $\nu$MSM, or implementing a conformal symmetry solution, which requires an understanding of how to include gravity in such a theory. This is something we cannot do at present.

\subsection{$\sgt$ Inflation}
Although the Higgs inflation of the $\nu$MSM has not been ruled out, it relies on the questionable assumption that new strong dynamics appearing at the scale of perturbative unitarity breakdown, $\Mp/\xi$, preserve the intact shape of the Higgs potential even above $\Mp/\xi$~\cite{Deg12}. We now consider another model of the scalar sector for the $\nu$NMSM, given in~\cite{Sha06} and developed further in~\cite{Ani09,Bez10}, in which the Higgs singlet $\sgt$ can be a light inflaton ($m_\sgt < m_h$) and thus provide an alternative to Higgs inflation.
The scalar potential of this model is
\begin{equation}
\label{eq:V2}
V = \lambda \paren{\Higgs^\dag \Higgs - \frac{\alpha}{\lambda}\sgt^\dag \sgt}^2 + \frac{\beta}{4}\paren{\sgt^\dag \sgt}^2 - \frac{1}{2}m^2 \sgt^\dag \sgt,
\end{equation}
where it is assumed that $m \ll \sqrt{\beta} \Mp$ so that chaotic inflation proceeds via the quartic term and, in contrast to~\cite{Sha06,Ani09,Bez10}, we require $\sgt$ to be complex to explain the hierarchical parameters of the $\nu$NMSM with an underlying symmetry. The potential \eqref{eq:V2} is then the most general renormalizable potential that respects a global U(1) symmetry under which only $\sgt$ is charged, assuming the bare mass term for the Higgs is negligible.\footnote{For a real $\sgt$, \eqref{eq:V2} was originally presented as the most general scale-invariant potential in which the scale invariance is explicitly broken by a mass term for $\sgt$~\cite{Sha06}.} Successful chaotic inflation requires $\beta \simeq 1.5 \times 10^{-13}$ to give the correct amplitude of adiabatic scalar perturbations and $\alpha \lesssim 10^{-7}$, $\lambda_{IJ} \lesssim 1.5 \times 10^{-3}$ to ensure that the flatness of the potential is not spoiled by radiative corrections from the loops of SM particles and sterile neutrinos~\cite{Bez10}.\footnote{As mentioned in~\cite{Sha06}, chaotic inflation with a quartic potential is disfavoured by WMAP data~\cite{Kom11}. However, only a very small non-minimal coupling to gravity of $\xi \gtrsim 0.0027$ can help bring this model in line with the data~\cite{Lin11}.} Achieving a sufficiently high reheating temperature $T_r > T_L$ for baryogenesis requires $\alpha \gtrsim 7 \times 10^{-10}$~\cite{Ani09}. Moreover, a value of $\lambda \simeq 0.13$ is required for $m_h \simeq 125$~GeV. For these parameters, expanding the potential \eqref{eq:V2} about its minimum gives the relations
\begin{equation}
\begin{gathered}
\label{eq:inflation}
\left< \Higgs \right> = \frac{v}{\sqrt{2}}, \quad \left< \sgt \right> = \sqrt{\frac{\lambda}{2\alpha}} v, \quad m_h \simeq \sqrt{2\lambda}v,\\
m_\sgt \simeq m \simeq \sqrt{\frac{\beta \lambda}{2 \alpha}} v, \quad \theta_{h \sgt} \simeq \sqrt{\frac{\alpha}{\lambda}},
\end{gathered}
\end{equation}
where $v = 246$~GeV. 
The upper bound on $\alpha$ can be further strengthened by limits on axion searches in the CHARM experiment~\cite{Bez10}. The mass range allowed by this experiment, $270\text{~MeV} \lesssim m_\sgt \lesssim 1.8\text{~GeV}$, corresponds to $2 \times 10^{-10} \lesssim \alpha \lesssim 8 \times 10^{-9}$ for $m_h \simeq 125$~GeV.
Note that we do not provide an explanation for the small values of $\alpha$ and $\beta$ in the scalar potential; we simply take their values to be within the range allowed by successful inflation. Also note that, for $\alpha \lesssim 8 \times 10^{-9}$, the invisible branching ratio of the Higgs is negligible.

Now consider the flavour sector of the $\nu$NMSM for this model of the scalar sector. As in~\cite{Sha06}, we assume an inflaton mass $m_\sgt \gtrsim 300\text{~MeV}$ so that the mixing angle $\theta_{h \sgt}$ is large enough to keep $\sgt^m$ in thermal equilibrium down to temperatures $T \ll m_\sgt$ via the interactions $\sgt^m \leftrightarrow e^- e^+,\sgt^m \leftrightarrow \mu^- \mu^+$, etc. We are therefore in the dark matter production case $r_f \gg 1$.
The ratio $m_\sgt / \left<\sgt\right> = \sqrt{\beta}$ is fixed by \eqref{eq:inflation} and \eqref{eq:lambda11} gives a value of
\begin{equation}
\label{eq:lambda11inflation}
\lambda_{11} \sim 3 \times 10^{-11}
\end{equation}
to produce correct dark matter abundance.\footnote{We have anticipated $S \simeq 1$ and $m_\sgt \simeq 400$~MeV in obtaining \eqref{eq:lambda11inflation}, though these parameters only have an $\mathcal{O}\paren{1}$ effect on $\lambda_{11}$.}
The absolute scale of $\left<\sgt\right>$, however, is not fixed. 
There is a relatively narrow window $7 \times 10^5\text{~GeV} \lesssim \left<\sgt\right> \lesssim 2 \times 10^6 \text{~GeV}$ that is consistent with the constraints on $\alpha$, the assumption $m_\sgt \gtrsim 300$~MeV, and the Lyman-$\alpha$ forest bound. Taking $\left< \sgt \right> \simeq 10^6$~GeV, a pattern of masses $M_{IJ}$ and couplings $\FaI$ that gives the correct dark matter abundance and baryon asymmetry can be achieved with two fields $\vartheta_1$, $\vartheta_2$, the values $\eta_1 \simeq 2 \times 10^{-3}$, $\eta_2 \simeq 5 \times 10^{-5}$, and the charge assignments given in table~\ref{table:FN2}.
\begin{table}[t]
\centering
\begin{tabular}{|c|ccccc|ccc|cc|cc|}
\hline
  & $N_1$ & $N_2$ & $N_3$ & $L_\alpha$ & $E_\alpha$ & $Q_i$ & $U_i$ & $D_i$ & $\Higgs$ & $\sgt$ & $\vartheta_1$ & $\vartheta_2$ \\
  \hline
  U(1) & 5 & -4 & -4 & -1 & -1 & 1/3 & 1/3 & 1/3 & 0 & 2 & 3 & 0 \\
  $Z_4$ & 0 & 1 & -1 & 0 & 0 & 0 & 0 & 0 & 0 & 0 & 0 & 1 \\
  $Z_2$ & 1 & 0 & 0 & 0 & 0 & 0 & 0 & 0 & 0 & 0 & 0 & 0 \\
  \hline
\end{tabular}
\caption{\label{table:FN2} Charge assignments for the $\sgt$ inflation scenario.
The global U(1) symmetry gives the structure \eqref{eq:lagrangian2} while the discrete $Z_4$ and $Z_2$ symmetries, together with the fields $\vartheta_1$ and $\vartheta_2$, give the required hierarchies in $F_{\alpha I}$ and $\lambda_{IJ}$.
}
\end{table}
Again, this is the simplest anomaly-free model we could find, though other charge assignments are possible.

Suppose this time that the Froggatt-Nielsen mechanism is used for generating the hierarchies in $F_{\alpha I}$ and $\lambda_{IJ}$, and hence $\eta_1 = \left< \vartheta_1 \right> / \Mp$ and $\eta_2 = \left< \vartheta_2 \right> / \Mp$. From table~\ref{table:FN2}, the Lagrangian for the flavour sector is then
\begin{equation}
\label{eq:lagrangianInflation}
\begin{aligned}
\Delta \mathcal{L} &= - f_{\alpha 2} \paren{\frac{\vartheta_1 \vartheta_2^\dag}{\Mp^2}} {\bar L_\alpha }{N_2}\Higgs - f_{\alpha 3} \paren{\frac{\vartheta_1 \vartheta_2}{\Mp^2}} {\bar L_\alpha }{N_3}\Higgs - \frac{h_{11}}{2} \paren{\frac{\vartheta_1^{\dag 4}}{\Mp^4}} \sgt \bar N_1^c N_1 \\
& \qquad - \frac{h_{22}}{2} \paren{\frac{\vartheta_1^2 \vartheta_2^{\dag2}}{\Mp^4}} \sgt \bar N_2^c N_2 - \frac{h_{23}}{2} \paren{\frac{\vartheta_1^2}{\Mp^2}} \sgt \bar N_2^c N_3 \\
& \qquad  - \frac{h_{32}}{2} \paren{\frac{\vartheta_1^2}{\Mp^2}} \sgt \bar N_3^c N_2 - \frac{h_{33}}{2} \paren{\frac{\vartheta_1^2 \vartheta_2^2}{\Mp^4}} \sgt \bar N_3^c N_3 + \text{h.c.},
\end{aligned}
\end{equation}
where $f_{\alpha I}$ and $h_{IJ}$ are $\mathcal{O}(1)$ couplings. Meanwhile, the scalar potential is given by
\eqref{eq:V2} plus the additional terms $\vartheta_i^\dag \vartheta_i H^\dag H$, $\vartheta_i^\dag \vartheta_i \sgt^\dag \sgt$, and $\vartheta_1^{\dag2} \sgt^3 +\vartheta_1^2 \sgt^{\dag3}$ involving $\vartheta_1$ and $\vartheta_2$. Again, we must assume that these additional terms in the scalar sector, which are allowed by the symmetries, have sufficiently small coefficients to preserve \eqref{eq:V2}. Once $\vartheta_1$ and $\vartheta_2$ acquire VEVs, \eqref{eq:lagrangianInflation} reduces to \eqref{eq:lagrangian2} with the textures
\begin{equation}
\label{eq:texture}
{F_{\alpha I}} \sim \left(
\begin{matrix}
  0 & \eta_1 \eta_2 & \eta_1 \eta_2 \\
  0 & \eta_1 \eta_2 & \eta_1 \eta_2 \\
  0 & \eta_1 \eta_2 & \eta_1 \eta_2
\end{matrix}
\right), \quad
{\lambda_{IJ}} \sim \left(
\begin{matrix}
  \eta_1^4 & 0 & 0 \\ 
  0 & \eta_1^2 \eta_2^2 & \eta_1^2 \\ 
  0 & \eta_1^2 & \eta_1^2 \eta_2^2 \\
\end{matrix}
\right).
\end{equation}
The parameters of the $\nu$NMSN are then
\begin{equation}
\begin{gathered}
F_{\alpha 2} \sim 1 \times 10^{-7}, \quad F_{\alpha 3} \sim 1 \times 10^{-7},\\
M_1 \sim 20\text{~keV}, \quad M \sim 4\text{~GeV}, \quad \Delta M \sim 10\text{~eV},
\end{gathered}
\end{equation}
up to $\mathcal{O}\paren{1}$ constants. As before, active neutrino mixing is anarchical (up to charged lepton corrections) and additional flavour symmetries must be used to produce hierarchies in the charged lepton and quark sectors. We also have the parameters
\begin{equation}
\alpha \simeq 4 \times 10^{-9}, \quad m_\sgt \simeq 400\text{~MeV}, \quad \theta_{h \sgt} \simeq 2 \times 10^{-4}.
\end{equation}
For these values, it can be shown that $\GB$ remains in thermal equilibrium roughly while $\sgt^m$ does (to temperatures below $m_\mu$) via the interactions $\sgt^m \leftrightarrow \GB \GB$. The near massless $\GB$ therefore contributes $\Delta N_\text{eff} \simeq 4/7$ to the effective number of neutrino species.

As a consistency check on this model, we must verify that $N_2^m$ and $N_3^m$ are out of thermal equilibrium at the characteristic temperature of leptogenesis $T_L \sim 2 \times 10^3$~GeV.
Since $m_\sgt < 2M$, the processes $\sgt^m \rightarrow N_2^m N_2^m$ and $\sgt^m \rightarrow N_3^m N_3^m$ are kinematically forbidden and the dominant processes are $H^\dag H \leftrightarrow N_2^m N_2^m$ and $H^\dag H \leftrightarrow N_3^m N_3^m$. These are out of equilibrium at $T_L$ for $\lambda_{23} \lesssim 0.01$, which is satisfied by \eqref{eq:texture}. One can also verify that the reheating temperature for $\alpha \simeq 4 \times 10^{-9}$ can be as large as $T_r \simeq 5 \times 10^3$~GeV~\cite{Ani09}, which is above the leptogenesis temperature.

This model demonstrates that, for a scenario in which $\sgt$ is a light inflaton, it is again possible to use symmetries broken at or near the Planck scale to obtain the pattern of Majorana masses and Yukawa couplings required for successful baryogenesis and dark matter production in the $\nu$NMSM. This model obeys all phenomenological constraints and provides an alternative to the Higgs inflation of the $\nu$MSM, but it requires small parameters in the scalar potential without explanation (a problem that plagues virtually all inflationary models) and does not improve the stability of the electroweak vacuum.

\section{Conclusion}
\label{sec:conc}

The $\nu$MSM is an extension of the SM that attempts to explain neutrino oscillations, dark matter, the baryon asymmetry of the universe, and inflation using only three sterile neutrinos with masses below the electroweak scale.
Despite the phenomenological successes of the $\nu$MSM, a further extension may be necessary to accommodate the Lyman-$\alpha$ forest bound, stabilize the electroweak vacuum, and allow for inflation.
In this paper, we have studied extensions of the $\nu$MSM by a Higgs singlet $\sgt$ that can address these issues and have demonstrated how the required pattern of masses and couplings in such models can arise from an underlying symmetry.

Our starting point has been an extension of the $\nu$MSM in which the decays of $\sgt$ give a primordial production of dark matter that is readily consistent with the Lyman-$\alpha$ forest bound and in which the VEV of $\sgt$ produces the Majorana masses of the sterile neutrinos.
For this next-to-minimal model, or $\nu$NMSM, we have considered two specific models of the scalar sector: one in which $\sgt$ helps stabilize the electroweak vacuum through a scalar threshold effect and one in which $\sgt$ is a light inflaton. For these definite examples, we have demonstrated that symmetries broken at or near the Planck scale can produce the required hierarchical pattern of Majorana masses and Yukawa couplings. The former model uses a $\text{U(1)} \times Z_3 \times Z_2$ symmetry while the latter uses a $\text{U(1)} \times Z_4 \times Z_2$ symmetry; both require a complex $\sgt$ rather than, as typically assumed, a real $\sgt$. We have not, however, provided an explanation for the parameters of the scalar sector or addressed the hierarchy problem associated with radiative corrections to the scalar sector.

The models presented in this paper satisfy all phenomenological constraints and make several experimental predictions that are distinct from the $\nu$MSM. These predictions include completely stable $N_1^m$ dark matter (hence no visible X/$\gamma$-ray signals from its decays) as well as anarchical active neutrino mixing angles (up to charged lepton corrections) with one active neutrino exactly massless.
Moreover, due to the complex $\sgt$, both models have $N_\text{eff} \simeq 3.6$ for the effective number of neutrino species while the former model has an invisible branching ratio of the Higgs of about 30\%. It will therefore be possible to test these models with the Planck experiment and the LHC in the near future.

\acknowledgments{I am grateful to Graham Ross for proposing this investigation and for much valuable input, as well as to Subir Sarkar for helpful discussions. This work was supported by the European Commission under the Marie Curie Initial Training Network UNILHC 237920 (Unification in the LHC era). Contents reflect only the author's views and not the views of the European Commission.}

\providecommand{\href}[2]{#2}\begingroup\raggedright\endgroup

\end{document}